\documentclass[%
reprint,
amsmath,amssymb,
aps,
pra,
]{revtex4-1}

\usepackage[T1]{fontenc}
\usepackage{hyperref}
\usepackage{graphicx,bm,xcolor,cleveref}

\newcommand{\ASlash}{\not\!\!A}
\newcommand{\kSlash}{\not\!\!k}

\begin{document}
	
	\title{Impact of ion mobility on nonlinear Compton scattering of an ultra-intense laser pulse in a plasma}
	\author{F. Mackenroth}
	\email{mafelix@pks.mpg.de}
	\affiliation{Max Planck Institute for the Physics of Complex Systems, 01187 Dresden, Germany}
	
	
	\begin{abstract}
We demonstrate that the nontrivial dispersion of a plasma driven by a high-intensity laser pulse qualitatively affects fundamental nonperturbative QED processes triggered by the laser pulse even in the case that no electrons remain in the interaction volume, e.g., due to ponderomotive expulsion. In the electron-free case this plasma effect is mediated by the response current of the residual ions as a dispersive effect on the laser propagation. The residual ions hence act as an effective background to the propagating electromagnetic laser field. We demonstrate that this has an impact on the fundamental nonlinear QED process of photon emission by an electron upon absorption of a large number of laser photons, called nonlinear Compton scattering. In two exemplary cases we find the ion plasma to suppress and enhance the photon emission rate by approximately $10\%$ for an intermediate and high laser intensity, respectively. The latter enhancement has no classical analog.
	\end{abstract}
	
	\maketitle
	
	\section{Introduction}
	
	The interaction of ultra-intense laser pulses with matter has received ever increasing attention over the past decade, due to tremendous technological progress, enabling modern facilities to deliver unprecedented electromagnetic energy densities in a controlled manner to small interaction volumes \cite{Hooker_etal_2006,Leemans_etal_2010,Zou_etal_2015,ELI_WhiteBook,CORELS}. The next generation of high-power laser facilities, becoming currently available for research around the globe, enables a plethora of groundbreaking fundamental physics studies \cite{DiPiazza_etal_2012} as well as numerous innovative technical applications, such as compact sources of high-energy radiation \cite{Corde_etal_2013}, relativistic particle beams \cite{Esarey_etal_2009,Daido_etal_2012,Macchi_etal_2013,Mackenroth_etal_2017a}, and damage-free laser pulse characterisation schemes facilitated by either analysing the radiation emitted from laser-scattered electrons \cite{Mackenroth_etal_2010,Har-Shemesh_etal_2012,Mackenroth_Holkundkar_2019}, or the electrons' own dynamics \cite{Mackenroth_Holkundkar_Schlenvoigt_2019}. All matter exposed to these ultra-high energy densities is immediately ionised to form a plasma of relativistic particles. And for a complete theoretical modelling of ultra-intense laser-plasma physics it is indispensable to identify and account for all physical effects potentially affecting the interaction. Notably, it has been pointed out that as soon as the laser photons transfer an average momentum to a charged particle which, in its rest frame, is on the order of its rest mass, quantum electrodynamics (QED) effects start to affect the interaction \cite{Ritus_1985}. Furthermore, due to the ultra-high photon densities of an ultra-strong laser, its photon field couples to the charged particle nonlinearly, making the use of nonperturbative field theoretical methods crucially important \cite{Reiss_1962}. These methods are capable of taking into account electromagnetic fields of arbitrary strength and their effects on charged particle dynamics by working in the Furry picture of quantum dynamics \cite{Furry_1951}. The two most common nonperturbative QED effects are the production of massive particle/anti-particle pairs, such as electron-positron pairs, as well as the emission of photons of such high energies that they exert a sizeable recoil on the emitting particle. This latter effect, in particular, labeled nonlinear Compton scattering, has been the scope of numerous studies over the past decade \cite{Harvey_etal_2009,Boca_Florescu_2009,Mackenroth_DiPiazza_2011,Krajewska_Kaminski_2012_a} and is conventionally modeled as the emission of a high-energy photon from a laser-dressed electron in vacuum. On the other hand, several applications explicitly rely on dense targets to control the laser propagation \cite{Stark_etal_2016,Gong_etal_2019} or nonperturbative QED effects themselves \cite{Wistisen_etal_2018,Wistisen_etal_2019}. For example, previously studied prolific X-ray sources from laser-driven electrons \cite{Chen_Maksimchuk_Umstadter_1998}, in all-optical setups \cite{Schwoerer_etal_2006,TaPhuoc_etal_2012,Sarri_etal_2014,Powers_etal_2014} can utilize the plasma as electron source \cite{TaPhuoc_etal_2003}. In this respect, even if the interaction is designed to be in vacuum, realistic ultra-intense laser setups involve many material components within the experimental chamber. Hence, applications, in any case, have to deal with a residual background gas pressure, quickly ionized to a plasma. Furthermore, in order to extract a sizeable detection signal, electrons enter the interaction volume as dense beams. Hence, there will be a background of charged particles, i.e., a plasma, even under best circumstances. In a classical framework, plasma effects on the radiation emission patterns of laser-driven electrons were found to be significant \cite{Castillo-Herrera_Johnston_1993}, just as could be expected from the impact of nonlinear effects on the emission of laser-driven electrons in vacuum \cite{Sarachik_Schappert_1970,Salamin_Faisal_1996}. This importance of plasma effects is fundamentally related to the collective response of a plasma to a strong laser-drive, which exhibits clear differences with respect to the vacuum response of single particles \cite{Akhiezer_Polovin_1956}. In addition, it is the scope of a large ongoing research effort to investigate how production of charged particles through nonperturbative QED processes can alter this plasma background or, at highest intensities, even create a plasma from previously empty space \cite{Bell_Kirk_2008,Gelfer_etal_2015,Grismayer_etal_2016,Jirka_etal_2017,Slade_Lowther_etal_2019}. At the core of large-scale numerical simulation schemes used to model such ultra-intense laser-plasma interactions lie nonperturbative QED rates obtained in vacuum. On the other hand, it was only recently demonstrated that a plasma background can significantly alter nonperturbative QED rates from their vacuum values, notably that of nonlinear Compton  \cite{Mackenroth_etal_2019}. Neglecting such alterations can have significant effects for the predictive power of established simulation schemes. It was thus deemed to be of highest relevance to accurately quantify potential effects of a nontrivial plasma background on fundamental nonperturbative QED effects and to provide estimates for which parameters the conventional vacuum rates need to be corrected. A fundamental problem of this task is how to model the plasma. Previously it was approximated as a homogeneous background composed of electrons and protons. However, an ultra-intense laser pulse induces strong envelope dynamics of the low-mass electrons, effectively pushing them out of the interaction volume through ponderomotive scattering \cite{Kruer_2003}. Hence, the actual electron density in the laser's path is not a priori known, and may even by vanishing for highest laser intensities. On the other hand, the heavier ions are less affected by a ponderomotive push and will hence stay in place longer, such that the interaction volume will still be filled with a charged plasma of heavy ions.
	\begin{figure}
		\centering
		\includegraphics[width=\linewidth]{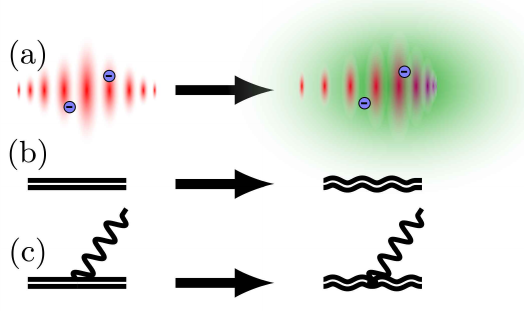}
		\caption{Schematic of the studied hierarchy of including plasma effects in first-principles QED calculations: (a) The classical high-intensity laser field experiences the plasma as a homogeneous background, affecting its propagation in the form of a nontrivial dispersion relation. (b) The altered laser properties affect the wave functions of a laser-driven particle, e.g., an electron. (c) Using the altered wave functions as basis set in nonperturbative QED calculations accounts for the plasma's dispersive effect.}
		\label{fig:Figure1}
	\end{figure}
	
	In this work, we estimate the effect of this remaining ion population on the laser propagation and consequently on nonperturbative QED effects triggered by the laser. We will follow the original approach of \cite{Mackenroth_etal_2019} and model the plasma background as a homogeneous charge cloud of number density $n$. Consequently, the plasma will only act by altering the laser's propagation dynamics through a nontrivial dispersion. On the other hand, even such a leading order approximation predicts significant alterations of the laser-dressed wave functions, used as basis sets in a Furry picture representation of QED. Hence, the emission characteristics computed from this basis set will be affected (s. \cref{fig:Figure1}). In accordance to earlier work \cite{Mackenroth_etal_2019}, the indirect effect of the background plasma on any nonperturbative QED effect involving electrons, mediated through the laser's altered dispersion, is modelled by replacing the vacuum wave functions of a laser-dressed electron (mass and charge $m_e$ and $e<0$, respectively) by their counterparts in a homogeneous plasma. The correspondingly required solutions of the Dirac equation have been intensely studied over the past decades (s. references in \cite{Mackenroth_etal_2019}).
	
	The paper is organised as follows: In the following \cref{sec:PlasmaProperties} we estimate typical time scales and plasma properties determining the dynamics of plasma heating and radiation emission. In \cref{sec:Dispersion} we introduce the model of the plasma dispersion used in this work. In \cref{sec:QED} we will reiterate the central aspects of performing nonperturbative QED calculations in a plasma-dressed laser field. And in \cref{sec:Numerics} we will present numerical simulations of the plasma-dependent alterations to the emitted energy. The final comments of \cref{sec:Conclusion} will then conclude this work. We will use natural units $\hbar=c=k_B=1$ unless explicitly stated otherwise.
	\section{Plasma model} \label{sec:PlasmaProperties}
	In this section we develop our model for the plasma response. A spatially inhomogeneous laser field $E_L(\bm{x})$, oscillating at a frequency $\omega_L$ exerts a ponderomotive force on any charged particle of mass $m$ and charge $q$ according to
	\begin{align}\label{eq:PonderomotiveForce}
	F_p = - \frac{q^2}{4m\omega_L^2} \nabla \left(E_L^2(\bm{x})\right).
	\end{align}
	Due to this force plasma particles are expelled from the interaction volume. Naturally, the ponderomotive expulsion will not be perfect. In order to estimate the maximal particle density the laser can expel from the interaction volume, we balance the ponderomotive force \cref{eq:PonderomotiveForce} with the charge separation field resulting from a number of $N$ charges being separated from the same number of opposite charges in the laser's focal volume by a distance $r$, which is simply given by 
	\begin{align}
	F_C = \frac{q^2 n V}{r^2},
	\end{align}
	where we estimate the particle number as the product of the particle density $n$ and the interaction volume $V$. The natural assumption for the interaction volume's size is that it is equal to the laser's focal volume $V\sim \pi w_0^2 l_R$, where $w_0$ is the laser's focal spot size and $l_R = \pi w_0^2/\lambda_L$ the Rayleigh length. Moreover, it is natural to assume that the particles will be transversely expelled from the focal volume to a distance of approximately $w_0$ and that the typical gradient length of the ponderomotive force in \cref{eq:PonderomotiveForce} will be of the same order. Then, the force balance of charge separation pull and ponderomotive push results in a maximal particle density which the laser can expel from its focal volume
	\begin{align}\label{eq:MaxDensity}
	n^\text{max} = \frac{E_L^2(\bm{x})}{2\pi m\omega_L^3 w_0^3}.
	\end{align}
	Particles in excess of this threshold remain inside the interaction volume. From the inverse dependence on the particle mass we conclude that electrons, being the lightest plasma particles, are expelled first and the heavier ions are much less affected by the ponderomotive push. Furthermore, the latter are much slower in response due to a smaller acceleration, as compared to electrons, resulting from an equal force. As a consequence, we can assume the ions to remain in place. We consequently model the plasma as a homogeneous background of charged ions only, which react to the laser field by forming a response current, which will affect the electromagnetic field propagation through the plasma. 
	
	This assumption of a homogeneous background ion plasma will be more reliable than in previously studies where an electron-proton plasma was assumed to be homogeneous \cite{Mackenroth_etal_2019}. In order to quantify the quality of the homogeneous assumption, we compare the characteristic time scales of the laser pulse's duration $\tau_L$ and the shortest collective response time of ions, given by the ion plasma period of protons
	\begin{align}
	\tau_{pi} = \frac{2\pi}{\omega_{pi}} \approx 477 \sqrt{\frac{\sqrt{1+\xi_i^2}}{n[10^{20}\text{cm}^{-3}]}} \text{ fs},
	\end{align}
	where we introduced the ion plasma frequency $\omega_{pi}=\sqrt{4\pi e^2 n/m_i^*}$ and assumed the ions to be protons with an effective mass relativistically increased through the laser pulse according to $m_i^* = m_p \sqrt{1+\xi_i^2}$, with $\xi_i = e E_L/m_p \omega_L$ the classical nonlinearity parameter of a laser with peak field $E_L$ for protons of mass $m_p$. Since ultra-intense laser pulses most commonly operate in ultra-short pulse mode $\tau_L\sim 10$ fs we see that $\tau_L\ll\tau_{pi}$, i.e., the laser pulse will pass through an almost unperturbed ion plasma, for densities $n\ll 2.3 \sqrt{1+\xi_i^2} \times 10^{22}\,\text{cm}^{-3}$. According to the ponderomotive heating model \cite{Piel_2014} the ions' temperature grows to $T_i \sim m_i^*$. Assuming that the collective ion motion builds up only over times scales $\tau_{pi}$, we can then estimate the laser-driven ion temperature to be $T_i \sim m_i^* \tau_L/\tau_{pi} \ll m_i^*$. Hence, the ion plasma can be approximated to be cold. Additionally, we are going to neglect ion collisions, as is common in ultra-intense laser-plasma interactions \cite{Macchi_2013}, as well as active plasma feedback such as the formation of instabilities or any plasmonic feedback. Furthermore, we are going to assume that the plasma's spatial extent is much larger than the ion plasma wavelength such that it is unperturbed by boundary effects, and that the ion plasma wavelength is much larger than the laser spot size $w_0 \ll 1/\omega_{pi}$. Under these assumptions, we can neglect the expansion of the surrounding ion plasma and treat its collective electrostatic field as a background to the motion of the laser-driven ions, permeating the laser-drilled plasma channel. Since the ions inside the laser channel will be driven from their equilibrium position inside this background field by the oscillating laser field only along the laser's polarisation direction, each will represent an oscillation dipole with respect to the unperturbed background field, and the laser's propagation through the ion plasma will be affected only by its collective response current. Naturally, a similar ionic response also affects the propagation of the emitted high-energy photons which we take into account here, in contrast to earlier work \cite{Mackenroth_etal_2019}, in order to also reliably model the emission of low-energy photons.
	
	Next, we compare the average time it takes a laser-driven electron to emit a photon to the typical time scale it takes the laser to heat the electrons to high temperatures. As we are considering the nonlinear regime $\xi\gtrsim1$, the time it takes for a photon to be emitted by laser-driven electron will be given by $\tau_\text{rad} \sim 1/(\alpha \omega_L \xi) \approx 10^2/(\omega_L[\text{eV}] \xi)$ fs \cite{Ritus_1985}. While we will find corrections to this vacuum estimate, it still remains of the same order of magnitude, for the parameters studied here. On the other hand, the heating time of the plasma's electron population is given by the characteristic time scale of a collective mode excitation, which is the inverse of the plasma frequency $\tau_{pe} \sim 2\pi/\omega_{pe} = (\pi \sqrt{1+\xi_e^2} m_e/(e^2 n))^{1/2}$, where $\xi_e = \xi_i m_p/m_e$ and we introduced the electron plasma frequency $\omega_{pe}=\sqrt{4\pi e^2 n/m_e^*}$ and estimated the electron's relativistic mass increase due to the oscillations in the laser field as $m_e^* = m_e\sqrt{1+\xi_e^2}$. From this consideration we conclude that it holds $\tau_{pe}=\tau_\text{rad}$ at the equilibrium plasma density 
	\begin{align}
	n = n^\text{eq} := \pi e^2 \sqrt{1+\xi_e^2} m_e \omega_L^2 \xi^2,
	\end{align}
	with $\tau_{pe} > \tau_\text{rad}$ for $n<n^\text{eq}$ and vice verse. In the following investigation we are going to consider both parameter regimes, such that the electron heating time is either significantly shorter or longer than the emission time, respectively. In the former case the electrons are first collectively heated to a high temperature by the laser, while their emission remains small. In the latter case, on the other hand, the electrons emit significant radiation already while being excited to a collective response by the laser. Nevertheless, in both cases the electrons are heated to high temperatures by the laser and pushed out of the interaction volume. It is worth pointing out, however, that even though we assume that the bulk of the plasma electrons will be expelled from the interaction volume by the laser well before the onset of photon emission, according to \cref{eq:MaxDensity} there will still be some residual electrons present in the laser's interaction volume. And since the Compton cross section is inversely proportional to a particle's mass, these electrons will radiate much more abundantly than the heavier ions. It is thus sensible to first consider the emission from a laser-driven electron here.
	\section{Ion plasma dispersion} \label{sec:Dispersion}
	To derive the ion plasma dispersion relation we start from the inhomogeneous Maxwell equations for the laser's electric and magnetic fields $\bm{E}_L$ and $\bm{B}_L$, respectively, which combine to the wave equation \cite{Jackson_1999}
	\begin{align} \label{eq:WaveEquation}
	\nabla \times \nabla \times \bm{E}_L + \frac{\partial^2 \bm{E}_L}{\partial t^2} &= - 4\pi \frac{\partial }{\partial t} \bm{j},
	\end{align}
	where $\bm{j}$ is the plasma current. In accordance with standard procedure, we will assume the vacuum solutions of this wave equation to be plane wave solutions, which are a complete basis of the electromagnetic field. They are of the form
	\begin{align*}
	\bm{E}_L &= \bm{E}_{L,0} \text{exp}\left[-i \left(\omega_L t - \bm{k}_L\bm{x}\right)\right]\\
	\bm{B}_L &= \bm{B}_{L,0} \text{exp}\left[-i \left(\omega_L t - \bm{k}_L\bm{x}\right)\right],
	\end{align*}
	where $\bm{k}_L$ is the laser's wave vector. With these solutions we can simplify the wave \cref{eq:WaveEquation} and and transform it to its Fourier space counterparts for the laser's Fourier component $\tilde{\bm{E}}_L$
	\begin{align}\label{eq:FT_WaveEquation}
	-\bm{k}_L (\bm{k}_L\tilde{\bm{E}}_L) + \bm{k}_L^2 \tilde{\bm{E}_L} - \omega_L^2 \tilde{\bm{E}}_L &= i 4\pi \omega_L  \tilde{\bm{j}},
	\end{align}
	where we also used the vector identity $\bm{k}_L\times \bm{k}_L \times \tilde{\bm{E}}_L = \bm{k}_L (\bm{k}_L\tilde{\bm{E}}_L) - \bm{k}_L^2 \tilde{\bm{E}}_L$. Under our assumptions introduced in \cref{sec:PlasmaProperties} the plasma polarisation will be a linear function of the driving laser field, such that the Fourier transform of the laser-driven current's Fourier transform will be given by the transform of the ions' equation of motion, which reads \cite{Piel_2014}
	\begin{align*}
	-i \omega_L m_i^* \tilde{\bm{v}} = - \frac{i\omega_L m_i^*}{n e} \tilde{\bm{j}} = e \tilde{\bm{E}}_L.
	\end{align*}
	Inserting the resulting charge current into \cref{eq:FT_WaveEquation} we arrive at the dispersion relation \cite{Piel_2014}
	\begin{align}\label{eq:FT_FieldWave}
	\left(\bm{k}_L^2 - \omega_L^2 + \frac{4\pi e^2 n}{m_i^*} \right) \tilde{\bm{E}}_L = 0,
	\end{align}
	where we additionally respected that for a transverse laser wave in a homogeneous plasma it will hold $\bm{k}_L\tilde{\bm{E}}_L\equiv0$. Naturally, for a nontrivial laser field \cref{eq:FT_FieldWave} is only satisfied for %
	\begin{align}\label{eq:Dispersion}
	\omega_L^2 = \bm{k}_L^2c^2 + \omega_{pi}^2,
	\end{align}
	where we found the ion plasma frequency $\omega_{pi}$ reappearing as the characteristic frequency of the ions' response current. This relation is the dispersion relation of a laser wave travelling through an ion plasma, which can alternatively be expressed as a refractive index for a photon of frequency $\omega$
	\begin{align}
	n(\omega) = \sqrt{1-\frac{\omega_{pi}^2}{\omega^2}},
	\end{align}
	which we are going to consider below.
	\section{QED model} \label{sec:QED}
	In order to estimate the effect of the background ion plasma on the emission of radiation from a laser-driven charge in a QED framework, we need to estimate the corresponding QED emission probability. As argued in \cref{sec:PlasmaProperties}, the emission from residual electrons can be stronger than that from the heavy ions, whence we are going to study the emission from such an electron, which we assume to remain in the ion plasma. We explicitly note that this does not contradict the assumption of near-complete electron expulsion from the laser channel, since the residual electron density derived in \cref{eq:MaxDensity} will be so small that it does not collectively affect the ionic plasma response. And, furthermore, in the surrounding plasma strong return currents will be driven, which will partly permeate through the interaction region \cite{Bell_etal_1997}.
	
	As introduced in \cref{sec:PlasmaProperties} we will model the laser pulse as a plane wave, depending on the space-time coordinates $x^\mu$ only through its invariant phase $\eta = x_\mu k_L^\mu$, where $k_L^\mu = \omega_L(1,n(\omega_L) \bm{\kappa}_L)$ is the laser's wave vector and $\bm{\kappa}_L$ its three-dimensional propagation direction, satisfying $\bm{\kappa}_L^2=1$. The laser's potential is then given by $A_L^\mu (\eta) = m_i\xi_i/\left|e\right| \epsilon_L^\mu \psi(\eta)$, where $\epsilon_L^\mu$ is the laser's polarisation and $\psi(\eta)$ its temporal shape, respectively. We then base the following discussion on the perturbative multiple-scale perturbation theory approach derived in \cite{Mackenroth_etal_2019}. In this perturbative framework, the Dirac equation and the solution ansatz are expanded in orders of $k_L^2/\omega_L^2 \ll1$ and the resulting power series solved iteratively. Such a perturbative approach is well suited to scattering at energy scales far above any binding barrier \cite{Heinzl_Ilderton_King_2016}. The solution of the Dirac equation for an electron of asymptotic momentum $p^\mu = \varepsilon(1,\beta \bm{n})$ in the presence of such a laser pulse propagating through a background plasma are derived to be
	\begin{align}
	\Psi_p(x) =& \bigg[\Phi_{V,p} + \frac{k_L^2}{2(k_Lp)} \delta \Phi_p \bigg]\text{e}^{-i (px) - i\Sigma_p(\eta)} \frac{u_p}{\sqrt{2 \varepsilon V}} \label{Eq:WaveFunction}\\
	\Sigma_p(\eta) =& \sigma_p(\eta) + \frac{k_L^2}{2(k_Lp)} \delta \sigma_p(\eta)  \nonumber \\
	\sigma_p(\eta) =&  e \frac{(pA_L)}{(k_Lp)} - \frac{e^2 A_L^2}{2(k_Lp)}\nonumber \\
	\Phi_{V,p} =& 1 + e \frac{\kSlash_L \ASlash_L (\eta)}{2(k_Lp)} \nonumber \\
	\delta \Phi_p =& \sigma_p(\eta)\left[1+e \frac{\kSlash_L \ASlash_L (\eta)}{(k_Lp)}\right] - \frac{e^2A_L^2(\eta)}{4(k_Lp)}\Phi_{V,p} \nonumber \\
	&- i e \frac{\kSlash_L \ASlash_L'(\eta)}{2(k_Lp)}. \nonumber
	\end{align}
	where $u_p$ is a vacuum four spinor, we used the slash notation $\not\!\!\!a = a_\mu \gamma^\mu$, with the Dirac matrices $\gamma^\mu$ and $a_\mu b^\mu = (ab)$ for any two four-vectors $a^\mu, b^\mu$. The wave function $\Psi_x(x)$ and its Dirac conjugate $\overline{\Psi}_p(\eta)=\Psi_p^\dagger (\eta)\gamma^0$ are then combined in the scattering matrix element of an electron of initial momentum $p_i^\mu$ to emit a single photon of momentum $k_f^\mu=\omega_f(1,n(\omega_f)\bm{\kappa}_f)$ and polarisation vector $\epsilon_f^\mu$ and change its momentum to $p_f^\mu$
	\begin{align}
	S_{fi} = \int d^4x \overline{\Psi}_{p_f} \not\!\epsilon_f^* \Psi_{p_i} \text{e}^{-i(p_i-k_f-p_f) - i\left(\Sigma_{p_i}(\eta)-\Sigma_{p_f}(\eta)\right)}. \label{eq:ScatteringMatrix}
	\end{align}
	From this scattering matrix element we obtain the emitted energy via the standard expression 
	\begin{align}\label{eq:EmittedEnergy}
	d E^\text{rad} = \omega_f \frac{d^3\bm{k}_fd^3\bm{p}_f}{(2\pi)^6} \sum_{\alpha,\beta} \left|S_{fi}\right|^2,
	\end{align}
	where $\alpha$ and $\beta$ represent the electron's and photon's final spin and polarisation degrees of freedom, respectively, and the total emitted energy $E^\text{rad}$ is obtained by integration over $d^3\bm{k}_fd^3\bm{p}_f$. Since the three space-time coordinates perpendicular to $\eta$ enter the integrand in \cref{eq:ScatteringMatrix} only linearly in the exponential phase, they translate to three energy-momentum conserving $\delta$-functions which collapse three of the final state momentum integrations. In order to simplify the remaining three-dimensional integral we note that in the regime $\xi\gtrsim 1$ an electron is driven to relativistic energies by the laser, whence it emits radiation only into a narrow cone around its instantaneous direction of propagation. Since a linearly polarised laser drives electron motion only in its plane of polarisation, we can confine our consideration to emission in that plane only. This corresponds to fixing the azimuthal angle to $\phi=0,\pi$ in a spherical coordinate system with the laser's propagation and polarisation direction as polar and azimuthal axis, respectively. Hence, in \cref{eq:EmittedEnergy} effectively there only remains a two-dimensional integration which we carry out numerically. 
	\section{Numerical studies and discussion} \label{sec:Numerics}
	\begin{figure}[t!]
		\centering
		\includegraphics[width=\linewidth]{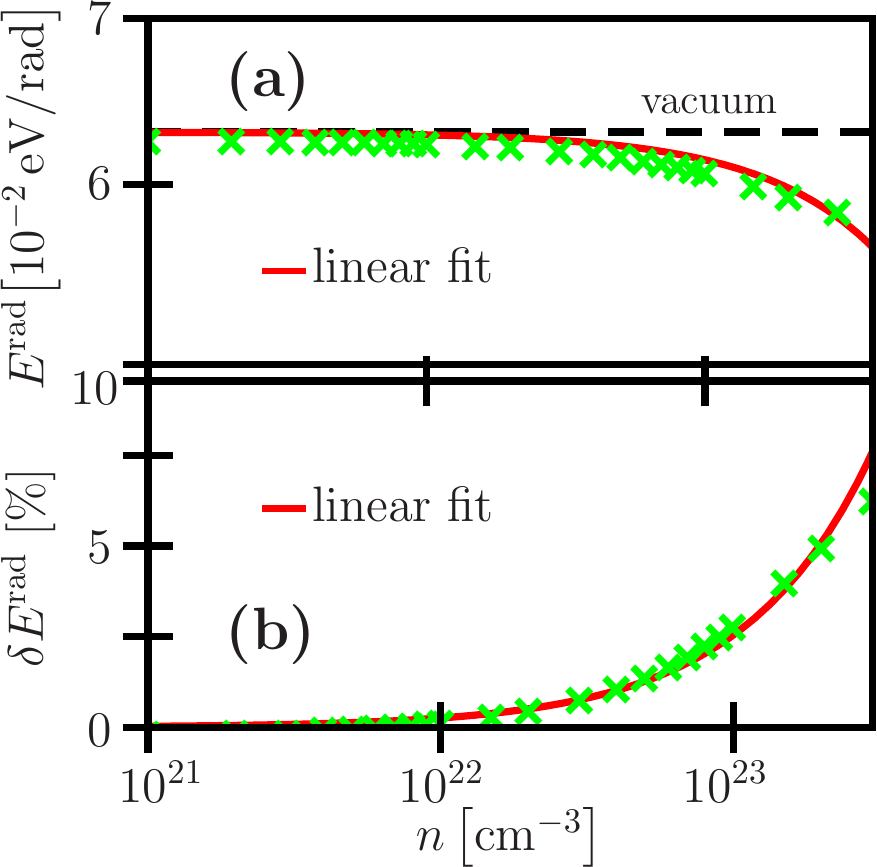}
		\caption{(a) Total emitted energy $E^\text{rad}$ of an electron driven by a laser pulse of intensity $I_L = 10^{19} \text{ W}/\text{cm}^2$ as a function of ion number density $n$. (b) Relative error of the exact result with respect to the vacuum result.}
		\label{fig:Figure2}
	\end{figure}
	
	In the following we consider an ultra-short laser pulse, modelled by the potential $A(\eta)=\sin^4(\eta/4)\sin(\eta)$, corresponding to a $\tau_L\approx 6$ fs pulse duration propagating through a homogeneous ion plasma of varying density $n$. We aim to model an optical laser field, whence we consider $\omega_L =1.55 $ eV, corresponding to a wavelength $\lambda_L=800$ nm. As mentioned above, we consider the simplest case of a plasma composed of protons with mass $m_p \approx 938$ MeV.
	
	We begin by studying the emission probability of a laser-driven electron inside a proton plasma bulk for a moderately relativistic laser intensity of $I_L=10^{19}$ W$/$cm$^2$ ($\xi_e \approx2$). In this case the maximal particle density compensating the laser's ponderomotive push from \cref{eq:MaxDensity} is approximately $n^\text{max}\approx7\times 10^{19}$ cm$^{-3}$, indicating that the proton density by orders of magnitude exceeds the residual electron density, which thus cannot be expected to perturb the proton dynamics. Furthermore, the equilibrium plasma density for the emission and heating times, derived above evaluates to $n^\text{eq} \approx 4\times 10^{19}$ cm$^{-3}$. As we are going to study much denser plasmas $n\gg n^\text{eq}$ below, we conclude that the electrons will be collectively heated to a high temperature before they emit radiation. We estimate the average thermal electron energy through the ponderomotive approximation $\varepsilon_\text{th} \approx m_e (1+\xi_e^2)^{1/2}\approx 2.2 m_e$ \cite{Wilks_etal_1992}, indicating that the electron will be only mildly relativistic. While the inclusion of the plasma's full thermal distribution function on the emission characteristics is certainly a relevant addition, it significantly complicates the numerical evaluation of the scattering matrix element, as a thermal average over an isotropic electron momentum distribution has to be performed. For the sake of tractability this is left to future work. We merely note that in the electron's rest frame the laser pulse will not be strongly differing from its lab frame properties and neglect its thermal motion to approximate the full result by the emission of an electron initially at rest $\varepsilon_i=m_e$. For this initial condition, we obtain a quantum nonlinearity parameter of $\chi := |eE_L|(k_Lp_i)/(m_e^3 \omega_L) \approx 6\times 10^{-6}$, which indicates quantum effects are unimportant. In fact, we find we find a reduction of the emitted energy $E^\text{rad}$ for an increasing plasma density (s. \cref{fig:Figure2}). The relative disagreement of the emitted energy energy with the vacuum result $E^\text{rad,vac}$ is well reproduced by a linear fit $\delta E^\text{rad} :=\left|E^\text{rad,vac} - E^\text{rad}\right| / E^\text{rad,vac} \approx 2.5\times 10^{-2}\, n[10^{23}\text{cm}^{-3}]$ which is qualitatively in agreement with earlier results obtained in a homogeneous, electron-ion plasma, in which the electrons are the carriers of the response current \cite{Mackenroth_etal_2019}. Furthermore, the negligible impact of quantum effects is highlighted by the fact that the observed reduction of the emitted energy with increasing plasma density is a classical effect due to the suppression of the formation of a radiating current by the plasma \cite{Mackenroth_etal_2019}.
	
	\begin{figure}[t]
		\centering
		\includegraphics[width=\linewidth]{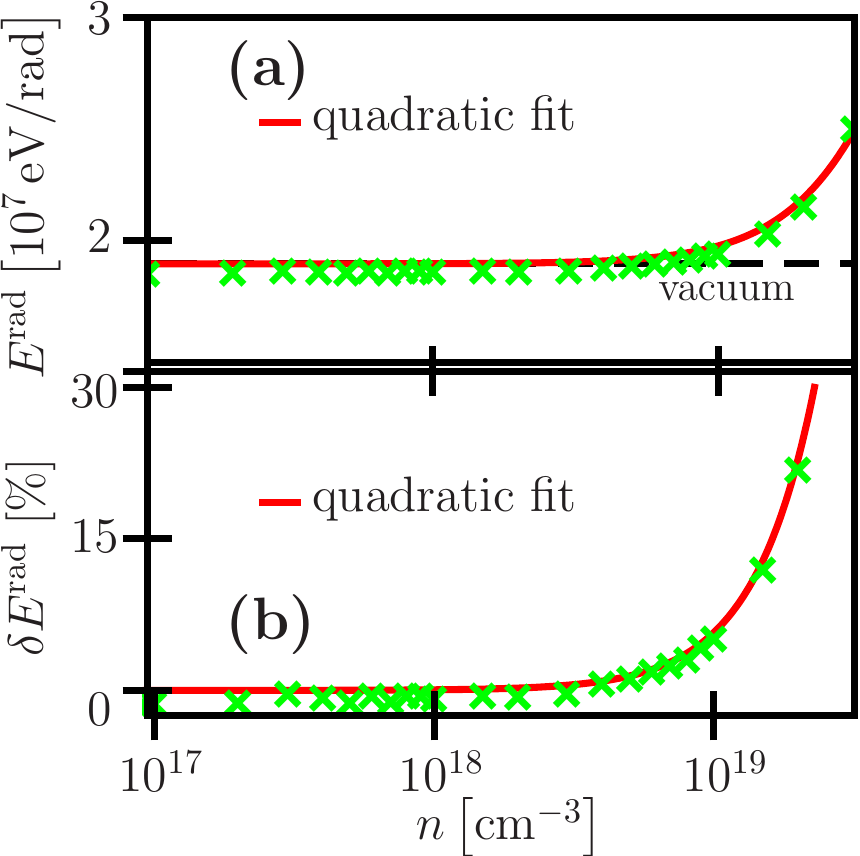}
		\caption{(a) Total emitted energy $E^\text{rad}$ of an electron driven by a laser pulse of intensity $I_L = 10^{22} \text{ W}/\text{cm}^2$ as a function of proton number density $n$. (b) Relative error of the exact result with respect to the vacuum result.}
		\label{fig:Figure3}
	\end{figure}
	
	Next, we turn to investigating the emission probability of a laser-driven electron inside a proton plasma bulk for a strongly relativistic laser intensity of $I_L=10^{22}$ W$/$cm$^2$ ($\xi_e \approx68$). In this case we find a significantly larger equilibrium plasma density $n^\text{eq} \approx 10^{24}$ cm$^{-3}$. Plasmas of such high density are opaque to optical radiation even when considering relativistic transparency and we find the emitted radiation to be corrected from the vacuum prediction beyond the applicability of the leading-order perturbation theory, developed here. Hence, we study plasmas of smaller density $n\ll n^\text{eq}$, whence we conclude the electron's collective response will be frozen on the time scales of photon emission. As a result, we can approximate even the plasma electrons to be cold and consider the emission of an electron of initial energy $\varepsilon_i=m_e$. We note that in this parameter set we obtain $\chi \approx 2\times 10^{-4}$, which indicates that the onset of quantum effects would not be expected in vacuum. Yet, even in a dilute proton plasma of density $n=10^{18}$ cm$^{-3}$ we find an enhancement of the emitted energy for an increasing plasma density (s. \cref{fig:Figure3} (a)). The relative disagreement of the emitted energy energy with the vacuum result is well reproduced by a quadratic fit $\delta E^\text{rad} :=\left|E^\text{rad,vac} - E^\text{rad}\right| / E^\text{rad,vac} \approx 5.7\times 10^{-2}\,  n^2[10^{19}\text{cm}^{-3}]$. And also in this case of a high-intensity laser pulse, the emission enhancement as well as the quadratic scaling are in qualitative agreement with results obtained for a neutral electron-ion plasma and not observed in a classical calculation of the emitted energy \cite{Mackenroth_etal_2019}. The appearance of these qualitatively non-classical signatures seems to indicate that the condition for the onset of nonlinear quantum effects may be altered by a background plasma.
	
	\section{Conclusion} \label{sec:Conclusion}
	
	We have demonstrated that the inclusion of the dispersive properties of a homogeneous plasma composed of heavy charged particles, such as ions, on a high-intensity laser pulse's dynamic characteristics, notably its dispersion, qualitatively alters the fundamental rate of radiation emission from a laser-driven particle, here shown on the example of photon emission from an electron embedded in a proton plasma. Quantitatively, we have shown, that for a mildly relativistic laser of intensity $I_L=10^{19}$ W/cm$^2$ in a proton plasma of density $n=10^{23}$ cm$^{-3}$, comparable to solid density, the emitted energy is reduced by several percent (s. \cref{fig:Figure2}). For a higher laser intensity of $I_L=10^{22}$ W/cm$^2$, on the other hand, we found that a proton plasma density of $n=10^{19}$ cm$^{-3}$ increases the emitted energy by almost $10\%$ (s. \cref{fig:Figure3}, in contrast to the classically expected suppression of emission. This result further corroborates the earlier finding, that the dispersive effect of a homogeneous plasma affects first-principles QED rate \cite{Mackenroth_etal_2019}. And this result shows that even in a high-power laser-driven plasma with all electrons ponderomotively expelled, there arises a sizeable correction to the vacuum rates of one of the most fundamental processes of nonperturbative QED, namely nonlinear Compton scattering. These results are relevant at current and upcoming high-power laser facilities.
	
	\section*{Acknowledgements}
	
	The author is thankful to Luis O.~Silva and Christoph H.~Keitel for interest in this work and fruitful discussions.

\end{document}